# Role of oxygen vacancies in Cr-doped $SrTiO_3$ for resistance-change memory

Dr. Markus Janousch[1*], Dr. Gerhard Ingmar Meijer[2*], Dr. Urs Staub[1*], Dr. Bernard Delley[1], Dr. Siegfried F. Karg[2] & Björn Pererik Andreasson[1]

[1]*Swiss Light Source, Paul Scherrer Institut, 5232 Villigen PSI, Switzerland*

[2]*IBM Research, Zurich Research Laboratory, 8803 Rüschlikon, Switzerland*

[*] These authors contributed equally to this work.

The future prosperity of information technology strongly depends on creating new device concepts with improved functionality and on successfully scaling of their characteristic lengths.[1] The spectrum of attractive novel non-volatile memory technologies currently being explored to sustain the increase of functionality in semiconductor devices ranges from magnetic random-access-memory[2, 3] and chalcogenide phase-change memory[4, 5] to resistance-change memory based on transition-metal-oxides.[6-8] The latter compounds can be conditioned such that they exhibit a bistable resistance state. The microscopic origin of the resistance-change memory in these transition-metal oxides is not understood. Here we investigate the relevance of oxygen vacancies for the resistance-change memory using the transition-metal oxide chromium-doped strontium titanate (Cr-doped $SrTiO_3$) as example. Laterally resolved micro-x-ray absorption spectroscopy and infrared thermal microscopy demonstrate that the conditioning process creates an electrically conducting path with a high density of oxygen vacancies which are localized at a Cr ion. Both resistance states exhibit metallic conduction. We propose that the microscopic origin of resistance switching in transition-metal-oxide-based memory is an oxygen-vacancy drift occurring in close proximity to one of the electrodes.

Dopants are of critical importance in semiconductor devices. Tiny amounts of dopants that act as donors or acceptors are introduced into the semiconductor crystal lattice to affect a significant change of the electronic properties of the semiconductor. This is particularly the case for the conventional complementary-metal-oxide-silicon (CMOS) technology, which is of fundamental importance for today's electronic devices. For these semiconductor devices, impurities have been characterized in great detail and their function has been fully elucidated.

In transition-metal oxides, defects are of similar importance. Defects are used, for example, to control the carrier doping responsible for the occurrence of high-$T_c$ superconductivity[9] and the colossal magnetoresistance effect.[10] In both cases, the doping can be achieved either by substituting cations of various valence states or by introducing oxygen ions or vacancies[11] into the crystal lattice.

Recently, it was shown that $SrTiO_3$ doped with Cr can be conditioned such that it exhibits a bistable resistance state.[6, 12, 13] Voltage pulses of opposite polarity switch the resistance of the perovskite reversibly between a high-resistance and a low-resistance state. These two different states persist after removal of the applied electrical bias. Cr-doped $SrTiO_3$ therefore holds the potential for nonvolatile memory applications. A similar resistance switching behaviour was found for other transition-metal oxides.[6-8] For oxygen-deficient single-crystalline $SrTiO_3$ it has been shown that the resistance of defect filament structures can be switched using a conducting tip of an atomic force microscope.[14],[15] Several models were put forward to explain the resistance-change memory in these transition-metal oxides. Tsui *et al.* proposed a conduction mechanism based on crystalline defects due to the applied electrical field.[16] Sawa *et al.* discussed an alteration of the Schottky barrier by trapped charge carriers in interface states.[17] Rozenberg *et al.* proposed a phase separation of non-percolative metallic domains.[18] More recently, a Mott metal-insulator transition at an interface due to band bending was discussed by Oka and Nagaosa.[19] However, in all these models the microscopic origin of the bistable resistance state in these transition-metal oxides remains unclear.

The memory devices reported here are based on single crystals of 0.2 mol% Cr-doped $SrTiO_3$. The as-prepared crystal exhibits a band-insulating character with a resistivity $\rho > 10^{11}$ Ω cm. The initially insulating Cr-doped $SrTiO_3$ becomes conducting after the crystal was exposed to an electrical field of $10^5$ V/cm for about 30 min, called conditioning process hereafter. Figure 1 (a) shows a current-voltage characteristics of a conditioned Cr-doped $SrTiO_3$ memory cell. The memory cell exhibits a hysteresis in the current-voltage characteristics, i.e., a bistable resistance state. Figure 1(b) displays the temperature dependence of the resistance of the memory cell in the low- and the high-resistance state. The decrease of the resistance upon cooling indicates that both resistance states are metallic. Moreover $R_{\mathrm{high}}(T) \propto R_{\mathrm{low}}(T)$, where $R_{\mathrm{high}}(T)$ and $R_{\mathrm{low}}(T)$ are the resistances of both states. This suggests that either the conductivity or the cross section of the metallic region changed upon resistance switching. Our experimental findings are in contrast to the models describing the resistance switching in terms of an alteration of a Schottky barrier[17] or non-percolative metallic domains,[18] which predict a non-metallic temperature dependence of the resistance

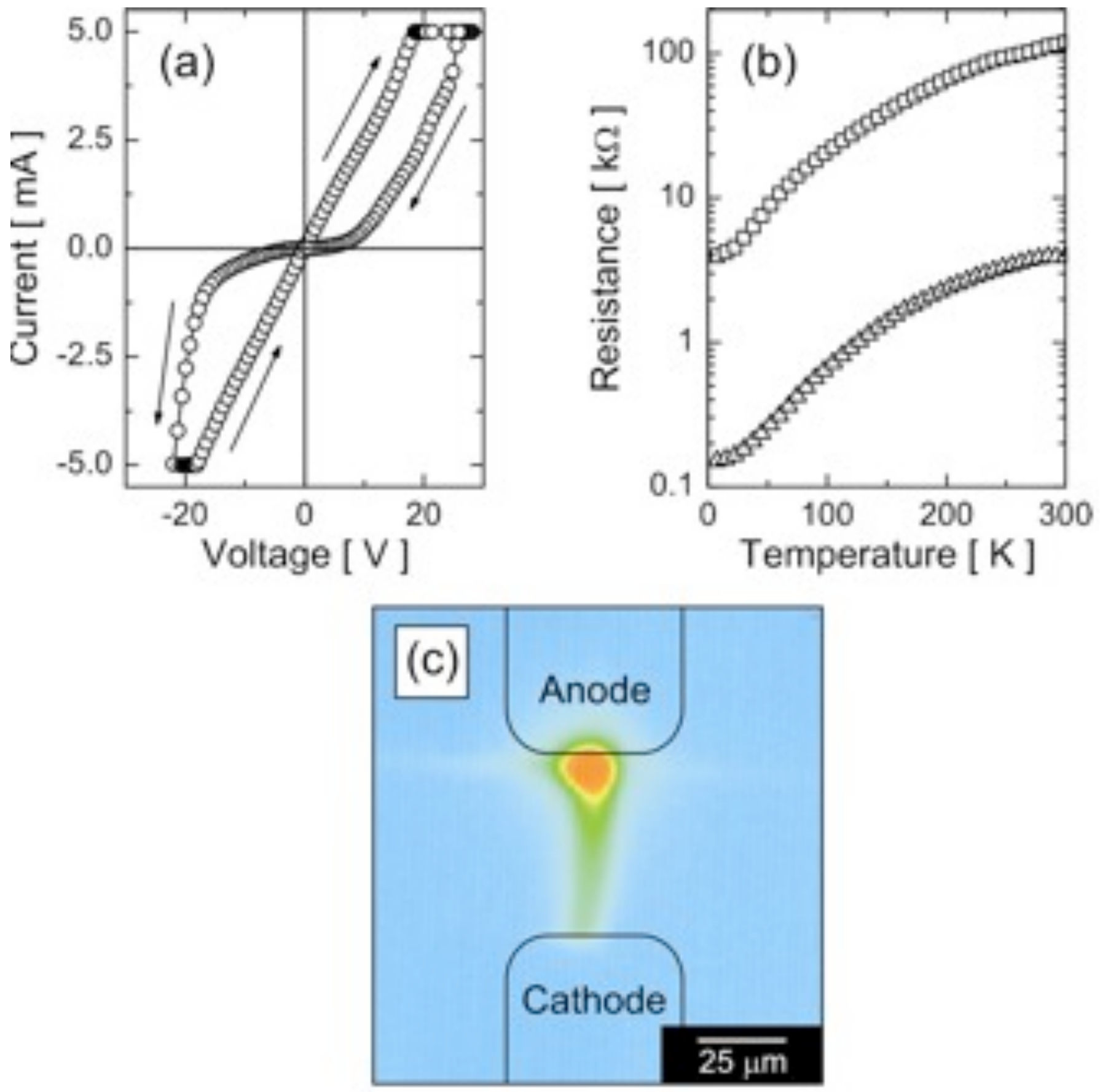

**Figure 1** Bistable resistance state and current path in a Cr-doped $SrTiO_3$ single crystal memory cell. (a) Current-voltage characteristics of the conditioned Cr-doped $SrTiO_3$ memory cell at ambient temperatures. (b) Temperature dependence of the resistance for the low- and the high-resistance state. (c) Infrared thermal micrograph of the memory cell with a current of +5 mA at a voltage of ~30 V applied. In the color scale, blue and red represent room temperature and elevated temperature, respectively. The electrodes used as anode and cathode for the conditioning process are indicated.

Figure 1(c) displays an infrared thermal micrograph of the memory cell that was collected while applying an electrical current of +5 mA at a bias voltage of ~30 V, i.e., approximately 150 mW power dissipated in the memory cell. The false-colour image reflects the temperature distribution of the memory cell. The temperature elevates in a laterally confined path between the electrodes. The majority of the power is dissipated near the anode electrode, reflected by the “hot spot.” This indicates that the local resistance is highest in the vicinity of this anode electrode. We did not attempt to obtain an absolute temperature calibration because the local temperature for a hot spot cannot be resolved by the microscope.

To obtain detailed information on the microscopic nature of the conducting path in Cr-doped $SrTiO_3$ memory cells, micro-x-ray fluorescence (XRF) maps and x-ray absorption near-edge spectra (XANES) at the Cr and Ti absorption *K*-edges were collected with high lateral resolution. Figure 2(a) shows the Cr *K*-edge XANES underneath the anode (A) and cathode (C). The reference spectrum (R) was taken 200 μm away from the electrodes and represents non-conditioned, i.e., undisturbed, Cr-doped $SrTiO_3$. The Cr *K*-edge XANES underneath the anode resembles $Cr^{4+}$, whereas the XANES at the cathode and reference positions is well represented by $Cr^{3+}$, as previously reported.[20] Also included in Fig. 2(a) is the difference between Cr *K*-edge spectra taken at the electrodes and the reference position. This contrast has a maximum at 6007.3 eV for Cr underneath the anode.

Figure 2(b) displays the Cr *K*-edge XANES taken in the conducting path near the electrode-to-crystal interfaces at the anode and the cathode (PA) and (PC), respectively. Both spectra exhibit essentially identical main-edge features and almost no energy shift compared with the *K*-edge of $Cr^{3+}$. However, for energies below the main edge, i.e., in the pre-edge regime, a pronounced increase of absorption is found. This *K* pre-edge region of transition metal ions is most sensitive to structural distortions. Also shown in Fig. 2(b) is the contrast of the Cr *K*-edge in the conducting path. Both (PA) and (PC) have a maximum in the pre-edge at 6004.3 eV. Moreover, XANES spectra of the Ti *K*-edge taken at the same positions do not show this contrast, they are identical within experimental accuracy.

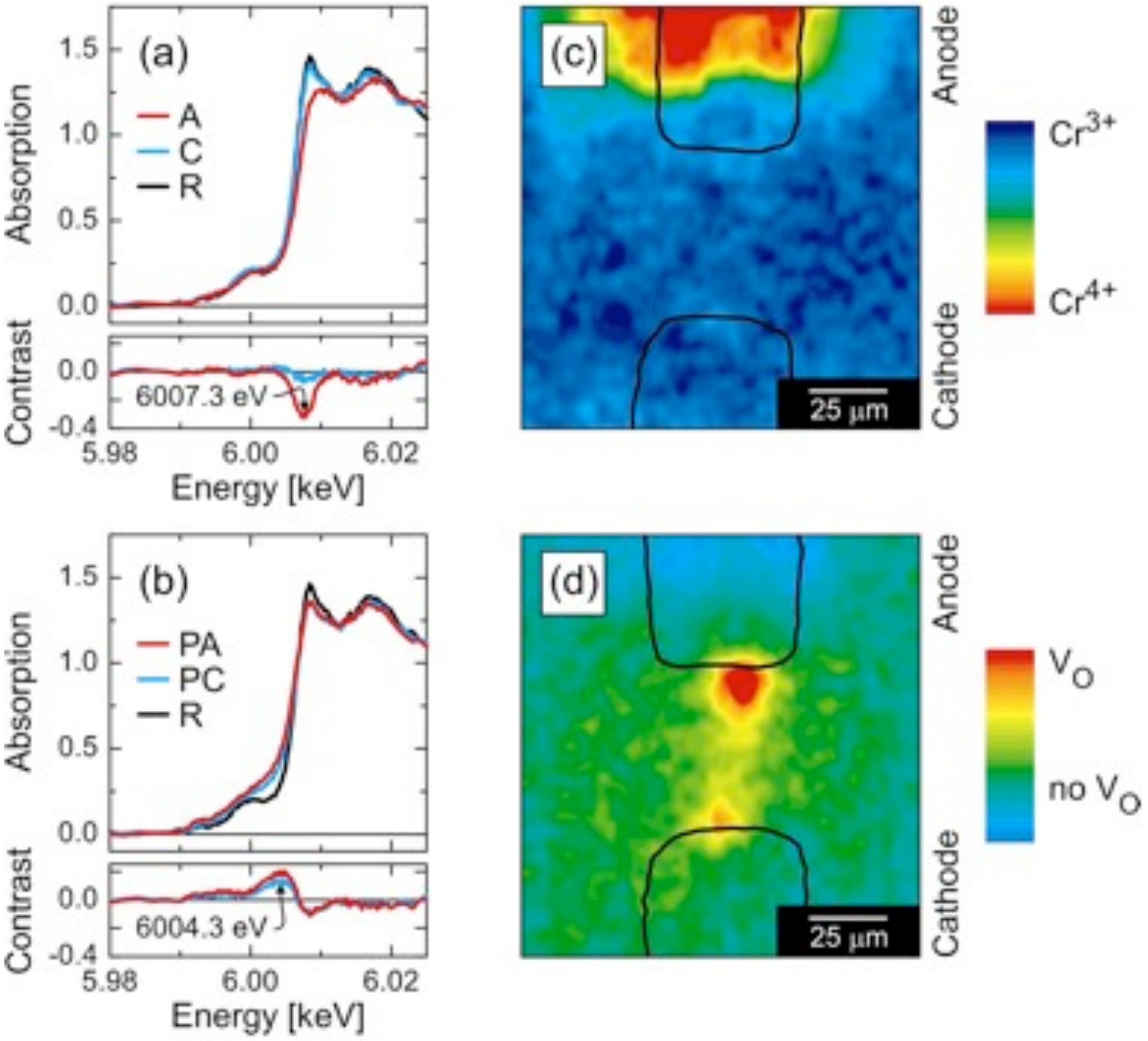

**Figure 2** Electronic states of Cr in a Cr-doped $SrTiO_3$ single-crystal memory cell. (a) Cr K-edge XANES underneath the anode (A) and cathode (C). The reference spectrum (R) represents non-conditioned Cr-doped $SrTiO_3$. Contrast is the difference spectrum (A) – (R) and (C) – (R). (b) Cr K-edge XANES in the conducting path near the anode (PA) and cathode (PC). (R) is the reference spectrum. Contrast is the difference spectrum (PA) – (R) and (PC) – (R). (c) Cr x-ray fluorescence map taken at 6007.3 eV for maximum Cr-valence contrast. In the colour scale, blue and red represent $Cr^{3+}$ and $Cr^{4+}$, respectively. (d) Cr x-ray fluorescence map taken at 6004.3 eV for maximum contrast at the Cr pre-edge region. In the colour scale, red represents oxygen vacancies $V_O$ in the Cr octahedra.

To characterize the lateral distribution of the electronic states of the Cr ions, XRF-maps were taken in the vicinity of the electrodes. Figures 2(c) and 2(d) show the Cr XRF-maps taken at 6007.3 and 6004.3 eV, respectively. The Cr fluorescence was normalized with the Ti fluorescence to correct for the absorption of the 100-nm-thick Pt electrodes. The map taken at 6007.3 eV, displayed in Figure 2(c), has maximum contrast between $Cr^{3+}$ and $Cr^{4+}$. In the colour scheme, blue and red represent a Cr valence of 3+ and 4+, respectively. Only in the far-field part of the anode is $Cr^{4+}$ found.[20] The Cr valence is 3+ underneath the cathode and also between the electrodes of the memory cell. It therefore seems doubtful that the Cr valence change is relevant for the resistance switching of Cr-doped $SrTiO_3$ memory cells.

Figure 2(d) is the Cr XRF-map taken at 6004.3 eV in the pre-edge region of the spectrum. For this energy a pronounced contrast, indicative for structural distortions, is found for the conducting path connecting the electrodes of the memory cell. The origin of this pronounced absorption increase at the Cr pre-edge becomes apparent if a comparison is made with a further reference crystal labelled (SR) in which a significant amount of oxygen vacancies were introduced using a post-anneal at 1400°C in an $Ar/H_2$ atmosphere followed by quenching to room temperature [Figure 3(a)]. The spectrum (PA) can be excellently reproduced with a linear combination of the spectra (SR) and (R). Since XANES probes the local structure of the absorbing atom this is strong evidence that the pronounced Cr pre-edge in the conducting path originates from oxygen vacancies $V_O$, located at octahedra surrounding the Cr ions.

This interpretation of the Cr XANES in the conduction path is further supported by density-functional-theory (DFT) band-structure ($DMol^3$)[21] calculations. For the DFT calculations supercells of $3 \times 3 \times 3$ $SrTiO_3$ unitcells, containing 1 Cr atom among the total of 135 atoms per supercell, were studied with $DMol^3$ DFT calculations. An extended variational basis set was used that includes 5*p* states for the Cr ion. Final integrations in momentum space were done with a 6 × 6 × 6 mesh. Figure 3(b) compares the density of Cr 4*p* states calculated by DFT for (i) a fully occupied lattice and (ii) a lattice with an oxygen vacancy $V_O$ in the octahedron surrounding the Cr atom. For scenario (ii), an increased density of states, and concomitantly an increased absorption, in the Cr pre-edge region are obtained, as observed experimentally. The calculations therefore indicate that oxygen vacancies are located in the first coordination shell of the Cr ion. The XANES in the conducting path cannot be reproduced with an oxygen vacancy in a higher Cr coordination shell, a Jahn–Teller distortion, an interstitial hydrogen atom, or a different valence state of the Cr ion.

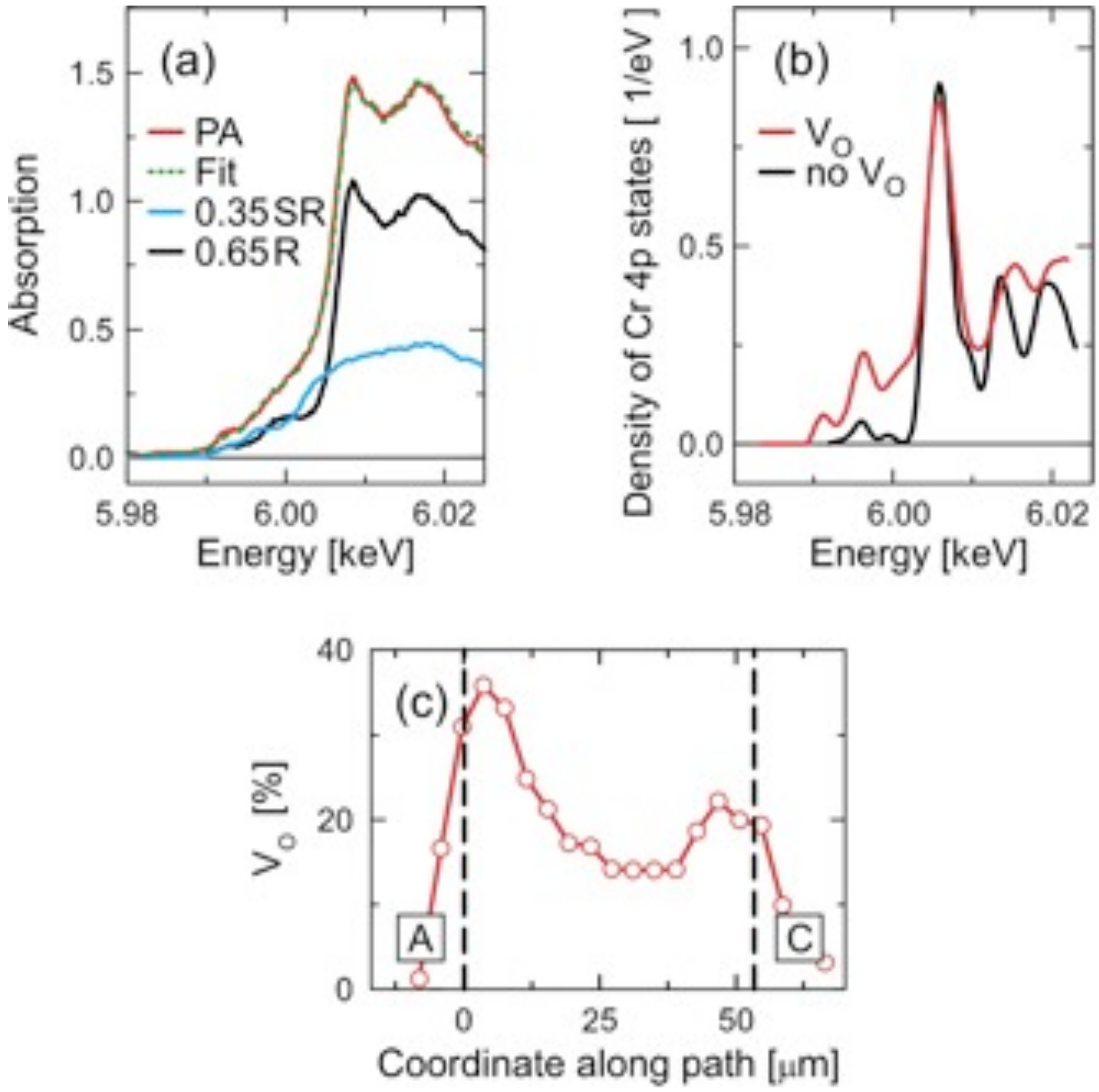

**Figure 3** Oxygen vacancies at the Cr ion in the conducting path in a Cr-doped $SrTiO_3$ single-crystal memory cell. (a) Cr K-edge XANES in the conducting path near the anode (PA). (R) is the reference spectrum. (SR) is the strongly reduced crystal. The dashed line is the linear combination of 65% (R) and 35% (SR). (b) DFT calculation of density of the Cr 4p states for (i) a fully occupied lattice and (ii) a lattice with an oxygen vacancy $V_O$ in the octahedron surrounding the Cr atom. (c) Oxygen-vacancy concentration at the Cr ion along the conducting path in the memory cell. The position of the anode and cathode are marked A and C, respectively.

To estimate the oxygen-vacancy concentration in the Cr-doped $SrTiO_3$ memory cell, we assume that for the strongly reduced and metallic reference crystal each Cr atom has one oxygen vacancy in its surrounding octahedron. In Figure 3(a) it can be seen that the XANES of the conducting path near the anode interface can be reproduced by admixing approximately one third of the strongly reduced spectrum to the reference spectrum (R). This indicates that, averaged over the sampling volume of $5 \times 5 \times 3.5\ \mu m^3$, approximately one third of the octahedra with a Cr ion in the path near the anode have bound an oxygen vacancy. Figure 3(c) shows the quantitative oxygen-vacancy concentration along the about 5-μm-wide path between the electrodes of the memory cell as extracted from the Cr XRF-map at 6004.3 eV. The oxygen-vacancy defect concentration is highest near the anode, with a second maximum at the cathode electrode.

The above-described micro-x-ray absorption spectroscopy and thermal microscopy now leads to the following picture for the resistance-change memory in Cr-doped $SrTiO_3$: The conditioning process introduced a few-micrometer-wide path of oxygen vacancies in the memory cell. These oxygen vacancies provide free carriers in the Ti 3*d* band leading to metallic conduction.[22] Cr plays the role of a seed for the oxygen vacancies, as concluded from the fact that these vacancies are preferentially introduced at the Cr ions. Since the oxygen vacancies are associated with the statistically homogeneously distributed Cr ions, the metallic conductivity occurs in a laterally extended area, i.e., the current path does not depend on the occurrence of dislocations as reported for undoped $SrTiO_3$.[15] After the conditioning process there remains an interface region near the anode that has a higher, though still metallic, resistance as revealed by the thermal microscopy. It is this high-resistance interface region of the memory cell that is exposed to the largest electrical field and "suffers" most power dissipation when a voltage pulse is applied to switch the resistance. These results lead us to the hypothesis that the resistance-switching process involves a drift of the oxygen vacancies along the applied bias field. For a negative bias applied to the "conditioning anode," oxygen vacancies, which are positively charged, are attracted into the high-resistance interface region, and the low-resistance state is obtained. If, on the other hand, a positive bias is applied, oxygen vacancies retract from the anode, and the memory cell switches back to the high-resistance state.

We have demonstrated that a high density of oxygen vacancies determines the path of electrical conduction in Cr-doped $SrTiO_3$ memory cells. The Cr acts as a seed for the localization of oxygen vacancies, leading to a statistically homogeneous distribution of charge carriers within the path. This warrants a controllable doping profile and improved device scaling down to nanoscale. The combination of laterally-resolved micro-x-ray absorption spectroscopy and thermal imaging allows us to conclude that the resistance switching in Cr-doped $SrTiO_3$ originates from an oxygen-vacancy drift to/from the electrode that was used as anode during the conditioning process. We propose that this oxygen vacancy concept is crucial for the entire class of transition-metal-oxide-based bipolar resistance-change memory.

## Experimenetal

*Synthesis of Cr-doped $SrTiO_3$ crystals:* The memory devices reported here are based on single crystals of 0.2 mol% Cr-doped $SrTiO_3$, grown in a $N_2/O_2$ atmosphere by floating-zone melting. The precursors were prepared using a stoichiometric mixture of SrO, $TiO_2$, and $Cr_2O_3$ sintered at 1350 °C in air. The crystals were slightly reduced by a post-annealing at 1150°C in an $Ar/H_2$ atmosphere for 6 h. Two 100-nm-thick Pt electrodes, $400 \times 50$ $\mu m^2$ in size, separated by a 50 μm gap, were patterned in planar geometry along the [010] axis on the [001] surface of the single crystals. As-prepared 0.2 mol% Cr-doped $SrTiO_3$ exhibits a band-insulating character with a resistivity $\rho > 10^{11}$ Ω cm.

*Conditioning of the crystals and measurements of physical properties:* For the conditioning process the Cr-doped $SrTiO_3$ was exposed to an electrical field of $10^5$ V/cm for about 30 min. A High Voltage Source Meter Unit, model K237 from Keithley Instruments Inc, applied the high voltage to the device. I-V curves were measured with the K237 and a Semiconductor Parameter Analyser, model 4155C from Agilent Technologies. The temperature dependence of the resistance in the crystal was measured with a Physical Property Measurement System from Quantum Design using a constant current of 10 μA at 30 Hz AC. Thermal images were taken with an infrared microscope equipped with a Hamamatsu IR camera.

*X-ray measurements:* Micro-x-ray fluorescence (XRF) maps and x-ray absorption near-edge spectra (XANES) were collected with high lateral resolution at the LUCIA beamline of the Swiss Light Source of the Paul Scherrer Institut.[23] The x-ray energy was chosen to be around the Cr and Ti absorption *K*-edges. The fluorescence of the Cr and Ti ions was collected simultaneously with a silicon drift detector. The beam spot-size at the sample was 5 × 5 μm$^2$. The sampling depth was approximately 3.5 μm. The data were recorded while the device was in the low-resistance state.

Acknowledgements

We thank J. G. Bednorz, W. Riess, and R. Allenspach for discussions, M. Schwarz and A. Jakubowicz, Bookham AG, Zurich, Switzerland for support with infrared microscopy, and U. Drechsler, F. Horst, R. Stutz, and D. Widmer for technical assistance. Part of this work was performed at the Swiss Light Source, Paul Scherrer Institut, Villigen, Switzerland.

Author Information

The authors declare that they have no competing financial interests.

Correspondence and requests for materials should be addressed to M.J. (markus.janousch@psi.ch) or G.I.M. (inm@zurich.ibm.com).